# A Voice Controlled E-Commerce Web Application

Mandeep Singh Kandhari
*School of Computing*
*Queen's University*
Kingston, Canada
mandeep.kandhari@queensu.ca

Farhana Zulkernine
*School of Computing*
*Queen's University*,
Kingston, Canada
farhana@cs.queensu.ca

Haruna Isah
*School of Computing*
*Queen's University*,
Kingston, Canada
isah@cs.queensu.ca

*Abstract*— **Automatic voice-controlled systems have changed the way humans interact with a computer. Voice or speech recognition systems allow a user to make a hands-free request to the computer, which in turn processes the request and serves the user with appropriate responses. After years of research and developments in machine learning and artificial intelligence, today voice-controlled technologies have become more efficient and are widely applied in many domains to enable and improve human-to-human and human-to-computer interactions. The state-of-the-art e-commerce applications with the help of web technologies offer interactive and user-friendly interfaces. However, there are some instances where people, especially with visual disabilities, are not able to fully experience the serviceability of such applications. A voice-controlled system embedded in a web application can enhance user experience and can provide voice as a means to control the functionality of e-commerce websites. In this paper, we propose a taxonomy of speech recognition systems (SRS) and present a voice-controlled commodity purchase e-commerce application using IBM Watson speech-to-text to demonstrate its usability. The prototype can be extended to other application scenarios such as government service kiosks and enable analytics of the converted text data for scenarios such as medical diagnosis at the clinics.**

## I. Introduction

Voice recognition is used interchangeably with speech recognition, however, voice recognition is primarily the task of determining the identity of a speaker rather than the content of the speaker's speech [1]. Speech recognition is a process of converting the sound of words or phrases spoken by humans into electrical signals to which a meaning is assigned [2] by comparing the signals with sets of phonemic representations for a close match. The phonemic representations are matched against words that are predefined in a word vocabulary. The goal of speech recognition is to enable people to communicate more naturally and effectively. This often requires deep integration with many natural language processing (NLP) components. Speech recognition can be applied to many domains and applications. It can remove barriers to human-human interactions by aiding people who speak different languages to be able to talk to each other without a human interpreter. It can be used in a messaging system to transcribe voice messages left by a caller into text that can be easily sent to the recipient through emails or instant messaging [3]. Speech recognition technology has made it possible to develop computer-based reading coaches that listen to students, assess the performances, and provide immediate customized feedbacks [1].

The traditional methods of data entry (keyboard and mouse) fail the accessibility requirements to support all types of users. Therefore, it is necessary to develop systems and applications with enhanced usability for all users. Our research focuses on a cloud-based Speech Recognition System (SRS) for e-commerce applications as a use case scenario. It is crucial for an organization or company to design and develop a web application that is informative, interactive and easily accessible to the users. Companies can use SRS to attract web audiences, advertise their products and services, educate customers about the features and usability of the products, and provide assistance in troubleshooting as are given through online chat programs today. Most of the existing applications lack the accessibility characteristics [4, 5], which poses a barrier especially to the visually-impaired users in terms of efficient access, and use of the information and web-services provided by the organizations. The web-accessibility standards established by the World Wide Web Consortium (W3C) Web Content Accessibility Guidelines (WCAG 2.1) [6] are not sufficient to address all accessibility problems [7]. SRS can enable people with physical and learning disabilities to interact with web applications at their own pace.

Over the years, various web assistive technologies like screen readers, special browsers, and screen magnification techniques have been developed to address the problems faced by visually impaired web users. These systems have helped the users either by reading, enabling voice commands, or providing ways for screen magnification to comprehend the contents on a web page. However, most of these solutions have failed in terms of accuracy as the frequency of misinterpretation is often high.

Web applications equipped with voice-enabled systems can not only provide flexibility in terms of users' choice of web interaction but can also increase the usability of the applications for the general users when they are unable to use the traditional human-computer interaction mechanisms. By allowing users to control the functionality of the applications with their voice, SRS can enhance users' browsing experience, and allow users to effectively convey their instructions and requests using natural languages.

In this paper, we present a study of the state-of-the-art speech recognition systems and propose a taxonomy of SRS. We also present a voice-controlled e-commerce application using IBM Watson speech-to-text service as a part of a comparative study with other speech-to-text systems such as Google and Amazon.

IBM Watson speech-to-text service uses advanced NLP and machine learning technologies for voice synthesization and text conversion. Our web-application takes a voice command, converts it to text, extracts meaning from the text, and then performs a wide variety of tasks including searching, browsing, reading and writing text.

The rest of the paper is organized as follows. Section II presents a background of the concepts, components, and applications of SRSs. The taxonomy of speech recognition systems is presented in Section III. Section IV presents a review of the related work including a comparison of some of the cutting edge SRSs. Section V describes an overview of the proposed framework while the implementation is described in section VI. Section VII concludes the paper with a discussion of the future work.

## II. BACKGROUND

The concept of interacting with a computer using voice has led to a new research area of SRS over the years which has significantly improved the human-to-computer interactions. It allows people to control their surroundings with their voice. Speech recognition technology can be used to dictate short messages, and thereby, reduce the effort needed to send a short text or email message, which otherwise needs to be typed. It can also be used to recognize and index speeches and lectures so that users can easily find the information that is interesting to them [1]. SRSs can be used to address the communication and interaction problems faced by people with disabilities.

Modern SRSs are built on statistical principles. The architecture of a typical SRSs, as shown in Fig. 1, consists of a voice input source or speech signal, feature extraction module, search engine, language model, acoustic model, adaptation model, and output or transcription components. The input data is the speaker's voice or speech, which is transformed into a speech waveform or signal data, and passed on to the feature extraction module for noise removal and transformation to the required representation format. The extracted signal is then moved to a search engine that is connected to language and acoustic models. The acoustic model includes knowledge about acoustics, phonetics, microphone and environment variability, and gender and dialect differences among speakers. The language model includes semantic knowledge or primarily meanings of words. It also describes what constitutes a possible word, what words are likely to co-occur, and in what sequence. It also houses the semantics and functions related to an operation a user may wish to perform. A modern SRS is able to handle a lot of uncertainties associated with speaker characteristics, rate, and style of speech; recognize basic speech segments, possible words, likely words, unknown words, and grammatical variations; process noise interference and non-native accents, and compute confidence scores of the results. This is the main function of the search engine component. The adaptation unit is used to modify the outputs from the acoustic or language models coming via the search engine to improve the overall performance [2].

There are many existing SRSs, and yet more are being developed. The next section presents the taxonomy for categorizing SRSs.

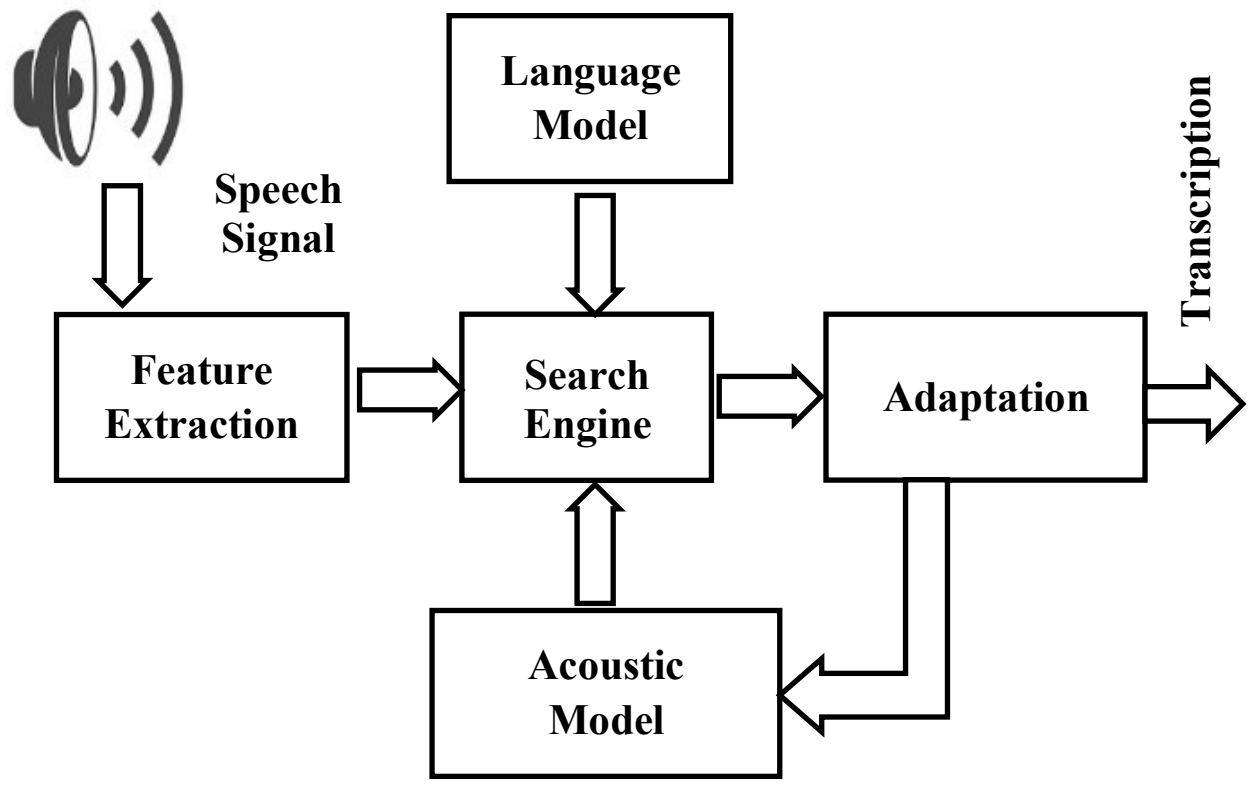

Fig. 1. The architecture of a speech recognition system.

## III. TAXONOMY OF SRS

Recent studies have shown that there are three major approaches to SRSs. These include acoustic phonetic, pattern recognition, and artificial intelligence approaches. In terms of design, SRSs can be divided into various categories on the basis of the type of input speech, vocabulary, style of speech, and speaker model [8]. Fig. 2 presents the taxonomy of SRSs. The criteria for the categorization of the SRS are described below.

### *A. Type of Speech*

*1) Isolated Words*: These models are only capable of detecting a single utterance at a time i.e. the words are isolated from the quiet phases at both the ends of the sample window. The system accepts a single word or an utterance and while the user is in the wait state, the system processes it and generates the appropriate result of the input speech. This approach computes changes in the energy of the input signal to leverage speech detection.

*2) Connected Words:* These systems allow users to give a set of words or utterances as inputs. The system identifies words with minimum pauses as units, that are trained in an isolated word mode. When the user enters the wait state, the system processes the input and generates the output accordingly.

*3) Continuous Speech:* In these systems, the user speaks continuously and it is the responsibility of the system to determine the word units and its segmentation. The implementation of such a system is quite difficult as special methods are required to recognize the text and to determine the utterance boundaries simultaneously.

*4) Spontaneous Speech:* These systems use advanced machine learning and NLP techniques to efficiently recognize spontaneous voice inputs, process and understand the meaning of the text, and generate appropriate responses. These types of systems are basically used in the query-based applications where the system returns answers to user queries.

### *B. Vocabulary*

The performance of an SRS depends on the size of the vocabulary it supports. The vocabulary size affects the accuracy, complexity, and the processing rate of a system. A system with a larger vocabulary size will have much better performance as compared to its peers since it allows the model to select from a wide range of words but will also require greater resources to maintain efficiency in terms of response time. SRSs can be further categorized based on the following vocabulary sizes that they support:

- Small - 1 to 100 words or sentences
- Medium - 101 to 1000 words or sentences
- Large - 1001 to 10,000 words or sentences
- Very-Large – More than 10,000 words or sentences

### *C. Style of Speech*

These SRSs apply the knowledge of accent and style of speech for word recognition. Although it requires knowledge about speech styles, style independent recognition is generally more challenging since, for matching phonemes, the internal representation of the speech must somehow be global.

### *D. Speaker Model*

Speaker independent recognition is more difficult because the internal representation of the speech must somehow be global enough to cover all types of voices and all possible ways of pronouncing words, and yet specific enough to discriminate between the various words of the vocabulary. For a speaker dependent system, the training is usually carried out by the user, but for applications such as large vocabulary dictation systems, this is too time-consuming for an individual user. In such cases, an intermediate technique known as speaker adaptation is used. Here, the system is bootstrapped with speaker-independent models which gradually adapts to the specific aspects of the user.

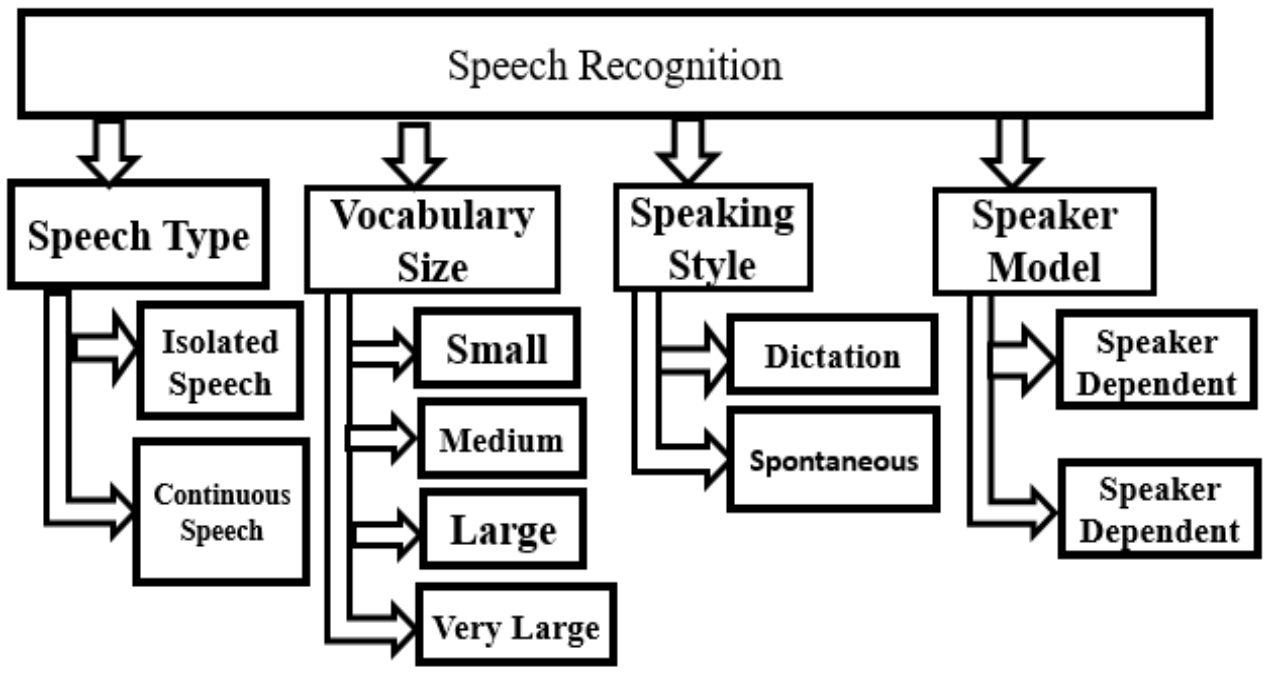

Fig. 2. Taxonomy of SRSs.

## IV. RELATED WORK

Many studies have been conducted where the researchers have integrated the SRS functionality with applications to make them usable by visually-impaired users. In an early study, Kevin Christian et al. [9] compared the performances of voice controlled and mouse controlled web browsing. Based on the subjective ratings of the 18 test subjects who used both controlling methods on hypertext forms, textual and numbering links concluded that the textual links are preferable to numbered links. Bajpei et al. [10] proposed a system where a SRS was incorporated in a browser. The system would accept users' voice as input and based on a predefined set of operations the browser would perform an operation.

The idea of the voice-operated browser was extended by Han et al. [11] such that it could be used in a smart TV environment. The research focused on navigating and controlling a dynamically generated hierarchical menu on a webpage with voice keywords. Sagar et al. [12] proposed an application where email services were combined with the text-to-speech and speech-to-text services thus enabling users to write, send, and read their e-mails. The main idea was to implement an add-on feature for the e-mail services.

### *A. Comparison of Cutting Edge SRSs*

In this section, we present a summary of comparative studies of different SRS from the literature. A number of offline and cloud-based SRSs are available in the market which differ in terms of the tradeoff between privacy and accuracy. The offline systems provide a high degree of privacy, which is preferred when working in a specific domain where security of the data is a big concern. However, they have lower accuracy compared to the less private systems due to the lack of access to knowledge resources imposed partially by the constraints on sharing information. Based on accessibility, SRS can be basically of two types, cloud-based and open source.

TABLE I. EXAMPLES OF VOICE RECOGNITION SYSTEMS (VRS)

| Cloud-based Proprietary SRS | Open source SRS |
|---|---|
| IBM Watson | Kaldi |
| Google Speech API | CMU Sphinx |
| Microsoft Speech API | HTK |

For comparing the quality of the SRS, a comparison[1] was conducted for cloud-based SRS, where a set of 3,000 different voice search phrases for e-Commerce were used. A recorded audio file with search phrases was given as an input to a number of SRSs and then the output text generated by these services were analyzed and compared with the actual search phrases. The quality of the results was determined by the phrase recognition rate (PRR) defined as the ratio of the number of recognized phrases (correct and incorrect) to all phrases as shown in Eq. 1, and the word error rate (WER) defined as the ratio of the number of correctly identified words to all phrases as shown in Eq. 2.

$$PRR = \frac{\text{Number of Words} - \text{Substitutions} - \text{Deletions} - \text{Insertions}}{\text{Number of Words}} \quad (1)$$

$$WER = \frac{\text{Substitutions} + \text{Deletions} + \text{Insertions}}{\text{Number of Words}} * 100 \quad (2)$$

Results of the experiments are given in Table II. In terms of quality, Google proved to be superior to the other systems as it was able to identify 73.3% of the text with only 15.8% WER and 73.3% PRR. Nuance's Dragon system achieved around 39.7% WER with a PRR of 44.1%. IBM Watson recognized 46.3% of the text with a WER of 42.3% and 46.3% PRR. The WERs for AT&T and WIT were similar, about 63.3%. However, the PRR

[a] http://blog-archive.griddynamics.com/2016/01/automatic-speech-recognition-services.html

for AT&T was better than that of WIT (32.8% as compared to 29.5%).

TABLE II. CLOUD-BASED SRS WER AND PRR

| Cloud-based SRS | WER (%) | PRR (%) |
|---|---|---|
| Google | 15.8 | 73.3 |
| Nuance | 39.7 | 44.1 |
| IBM | 42.3 | 46.3 |
| AT&T | 63.3 | 32.8 |
| WIT | 63.3 | 29.5 |

Këpuska and Bohouta [13] built a tool to compare the performances of the cloud-based Google and Microsoft Speech Recognition (SR) APIs with the open source CMU Sphinx system. A recorded file was given as an input to the systems and the system generated output was used for calculating the PRR and WER. The results concluded that the Google Speech API had a better WER performance of 9% followed by the Microsoft Speech API (18%) and Sphinx-4 (37%).

The performances of open-source speech recognition toolkits are evaluated by Gaida et al. [14] on two datasets namely the *vermobil1 corpus* and the *wall street journal1 corpus*. They compared the Hidden Markov Model Toolkit (HTK) with HDecode and Julius, CMU Sphinx, and the Kaldi toolkit. The authors reported that the Kaldi toolkit provided the most promising results followed by the Sphinx and HTK. The results generated by the Kaldi toolkit were better since a significant amount of knowledge and effort had been spent on training the model.

TABLE III. PERFORMANCE OF SENTENCE RECOGNITION [15]

| SRS | Kaldi | | Watson | |
|---|---|---|---|---|
| **Sentences with (total 141)** | **No. of Sentences** | **% error** | **No. of Sentences** | **% error** |
| **Errors** | 135 | 95.7% | 28 | 95.7% |
| **Substitutions** | 39 | 27.7% | 23 | 67.3% |
| **Deletions** | 15 | 10.6% | 21 | 10.6% |
| **Insertions** | 90 | 63.8% | 10 | 27.7% |

TABLE IV. PERFORMANCE OF WORD RECOGNITION [15]

| SRS | Kaldi | | Watson | |
|---|---|---|---|---|
| **Total Words: 2060** | **Number of words** | **% error** | **Number of words** | **% error** |
| **% total error** | 205 | 10% | 63 | 3.1% |
| **% correct** | 1855 | 90% | 1997 | 96.9% |
| **% substitution** | 51 | 2.5% | 22 | 1.1% |
| **% deletions** | 48 | 2.3% | 24 | 1.2% |
| **% insertions** | 106 | 5.1% | 17 | 0.8% |
| **% Word accuracy** | 84.9% | | 96.1% | |

Elma Gazetić [15] compared the performances of Kaldi and IBM Watson, using an audiobook 'Emma from Jane Austin' as an input. The performances were measured in terms of WA and WER. It was found that the cloud-based applications had far superior performances over the open source tools such as Kaldi because the former systems are trained for thousands of hours as compared to the latter which was trained only for a few hundred hours.

## V. OVERVIEW OF THE FRAMEWORK

Majed Alshamari [16] evaluated the accessibility of the three well known Saudi e-commerce websites with five testing tools: Achecker, TAW, Eval Access, Mauve, and FAE. The study concluded that the websites were not competent enough to address the basic accessibility problems such as navigation, readability, timing, and input assistance.

According to the McNair's report [17], the worldwide retail e-commerce business in the year 2018 was expected to reach $2.774 trillion dollars which accounts for 11.6% of the total retail sales. The sales numbers are expected to touch $4 trillion dollars' mark by the end of the year 2021, which constitute 15.5% of the total retail sales. The numbers are evident enough to depict that e-commerce has emerged as an alternative to the traditional retail business, giving the users a convenient option of ordering the products online. However, the accessibility issues [14] in the websites discourage a large number of visually impaired users who remain deprived of the facilities provided by the websites and the e-commerce business loses some of its potential customers. This motivated us to design a SRS based e-commerce system to address the accessibility problems faced by the disabled users. This section presents an overview of the design of the framework. Implementation details of the prototype are described in the next section.

### *A. Watson SRS for E-COMMERCE*

Poor accessibility of websites causes negative impacts on web users and causes hit scores for those websites to decrease with time. For e-commerce businesses, this can be disastrous as most of the transactions are done online. Researchers have discussed assistive technologies like screen readers, voice browsers and a variety of adaptive strategies that users apply to surf the internet, but these are not sufficient and often fail to address the accessibilities issues. Borodin et al. [18] describe the usability problems faced by the screen-reader users. Even though the technology allows access to web pages, it can often become unbearable especially if a large number of links are present on the page. The screen-reader would make announcements every time it encounters a link, thus making it impractical and annoying for the users. An e-commerce web application is a perfect example as it often contains a large number of links to different product pages. The incorporation of speech services can provide an alternative mechanism to the users where users can directly interact with the applications. SRS will further enhance the accessibility of a webpage and will empower the users with the ability to efficiently access, navigate, skim through a large amount of information, get input assistance, and even control the functionality of the webpage.

Based on the comparative study of the existing SRS in Section 3, we can conclude that the performances of cloud-based systems are better than the open source systems in terms of WER and PRR. Higher accuracy ensures that the users do not have to give the same voice commands repeatedly to the application. The open source tools can also support high accuracy rates, but a significant amount of time is needed to train and customize the system for a particular application. The main motivation for using Watson speech services is that access to the research platform and the software are provided by our collaborators from IBM. Also according to IBM tech report [19], the use of deep learning technologies has allowed Watson speech recognition to achieve 5.5% WER, which gives a good motivation to use Watson speech services in our research. Building a highly accurate SR engine requires access to knowledge about grammar and language structure along with mechanisms to compose audio signals and detect keywords. SRS should allow real-time transcription of the voice which is updated continuously with the incoming voice data. The custom model support can provide the freedom and flexibility to train the service according to the nature of the application. All these features are supported by Watson and make it a great tool to use in our web application.

The architecture of our prototype e-commerce web application using Watson SRS is depicted in Fig. 3. Users, with the help of the microphones in their computer systems, will give voice commands to the application. The input signal is then streamed through the Watson Speech-To-Text (STT) service which converts the audio signal to the appropriate text. One of the key features of the Watson STT service is the capability of efficiently identifying keywords in the text. Since every web application has to perform a predefined set of operations, the application can look for these keywords. When a matching keyword is found, the application can trigger a predefined functionality. In the case an application cannot find any matching keyword, it can use Watson Text-To-Speech (TTS) service to generate a voice response to inform the user about it.

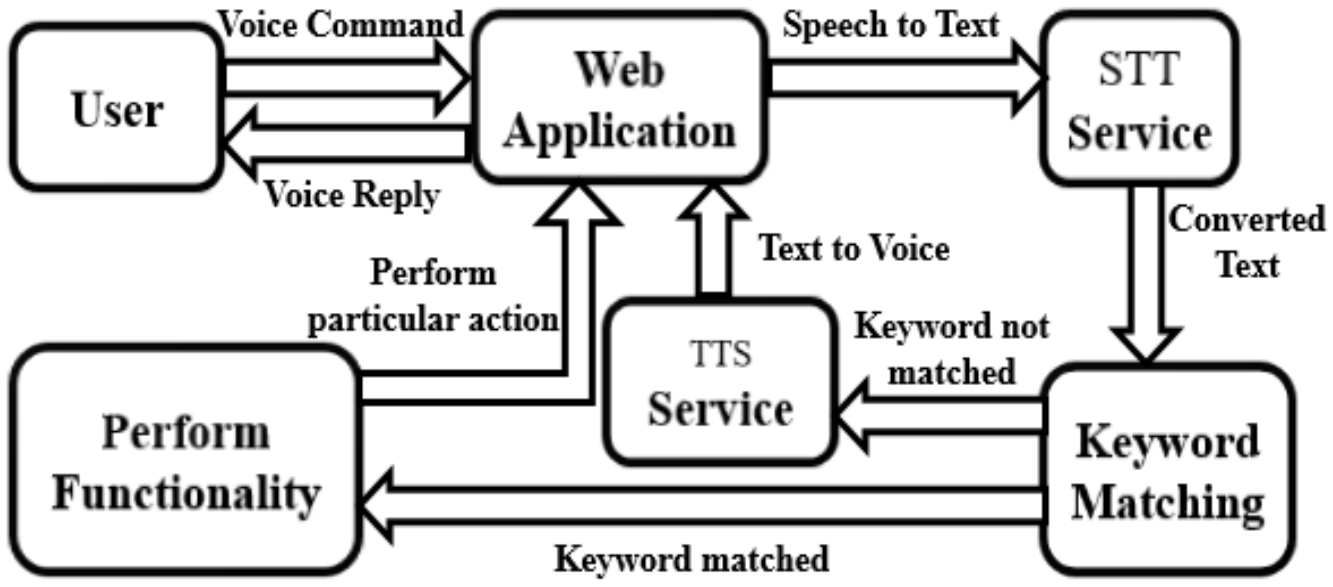

Fig. 3. The architecture of the Web Application.

The performance of a SRS depends on the size of the vocabulary it supports for a certain level of accuracy, complexity, and information processing rate. The vocabulary size of the cloud-based Watson system is very large and thus the STT and TTS generate good results. We implemented continuous speech type so that the user can make descriptive requests to the application without stopping. The other aspect of the SRS is that it is speaker independent i.e., the application can support voices and the pronunciation of words on a global level thus making it useful for a greater audience. Lastly, the SRS uses advance deep neural network techniques to generate accurate and precise voice transcriptions.

In our SRS application for e-commerce, we demonstrate a voice-controlled web-based commodity purchase scenario, which incorporates all the above aspects. The system accepts a user-query as a voice command, looks for particular keywords in it, and based on the closest match found, executes a job like searching for a particular item and traversing through different web pages. The supplementary information provided by the Watson services along with the transcribed words are used to build an intelligent query and matching system.

## VI. IMPLEMENTATION DETAILS

The architecture of our proposed e-commerce web application integrated with the Watson SRS is depicted in Fig. 3. In this section, we describe the implementation details of our prototype application.

### *A. Software Components*

We used Watson STT and TTS to build our SRS prototype e-commerce application. The application combines a) client-side scripts (JavaScript and HTML) used to present information to the users, b) a back-end application powered by server-side scripts (JavaScript and Python), and c) a database system, where the latter two allows storing and retrieving information and generating responses to send back to the client. For our application, the components were picked on the basis of the survey[2] where 20,000 developers were asked for their opinion about the tools and frameworks currently used by them. The results showed that the React JavaScript library and the Express are currently the most preferred tools. For the database, we chose MongoDB, which is a non-relational database that can store information in the BSON (Binary JavaScript Object Notation) format. It provides an efficient way of building a highly scalable application supporting diverse data types. Data about the merchandise which is used to build our e-commerce website was obtained from the edX “front-end capstone project” [3]course. The course teaches how to develop an e-commerce website and uses a dataset containing a list of products on their cloud in JSON format. This data is retrieved and fed into our application for rendering the products on the screen.

The different components and the information flow of our SRS e-commerce web application are shown in Fig. 4.

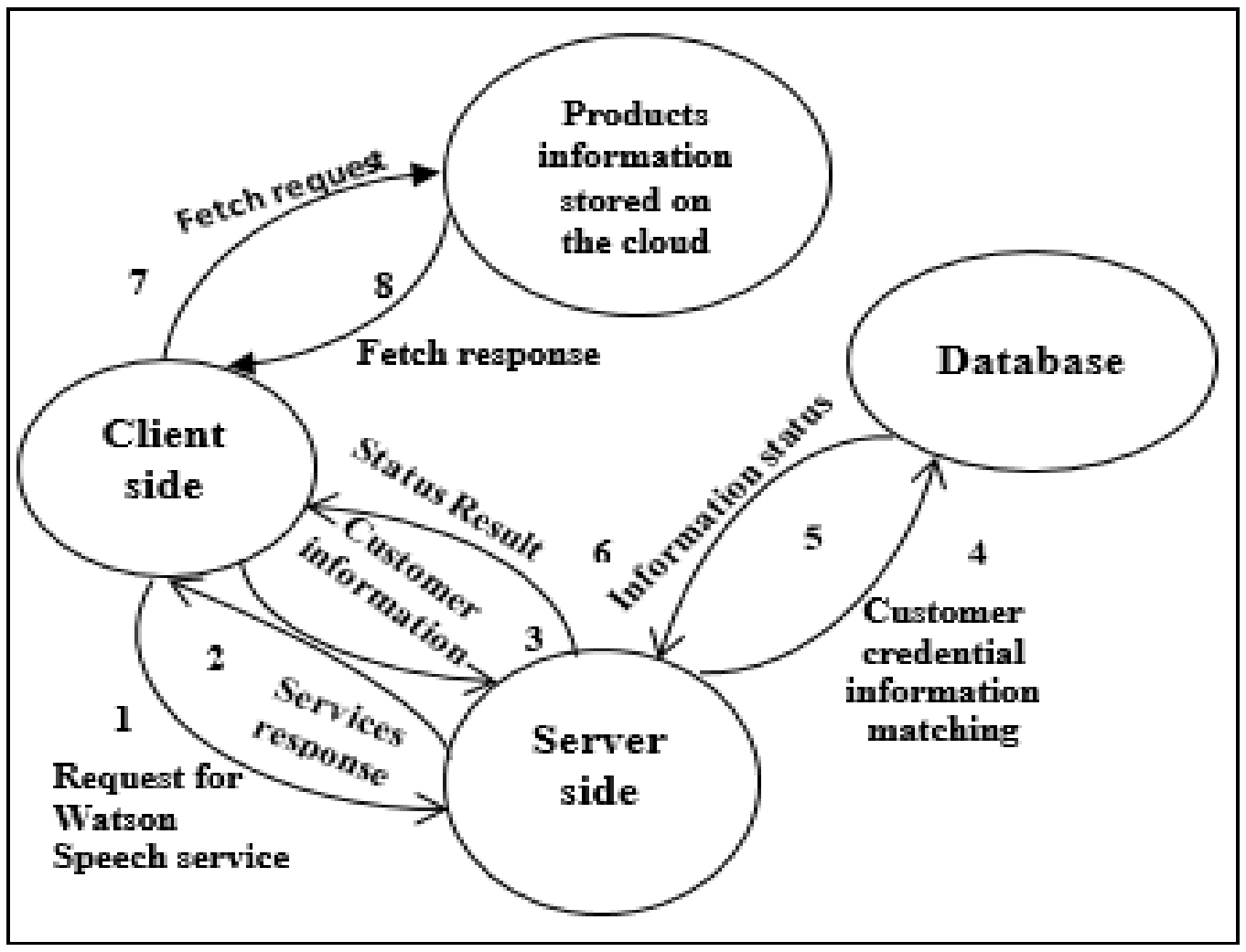

Fig. 4. Information flow in our SRS web application.

[b] https://stateofjs.com/

[c] https://courses.edx.org/courses/course-v1:Microsoft+DEV238x+2T2018/course/

1) *Information Flow:* When the application loads for the first time, the client script makes a request to the server for the IBM Watson STT and TTS services. The backend, after verifying the application's credentials, sends a success response back to the frontend enabling it to use the services as shown in Fig. 5.

```
const sttAuthService = new watson.AuthorizationV1(
  Object.assign(
    {
      username: '[redacted]',
      password: '[redacted]'
    },
    vcapServices.getCredentials('speech_to_text')
  )
);

app.use('/api/speech-to-text/token', (req, res)=> {
  sttAuthService.getToken(
    {
      url: watson.SpeechToTextV1.URL
    },
    function(err, token) {
      if (err) {
        console.log('Error retrieving token: ', err);
        res.status(500).send('Error retrieving token');
        return;
      }
      res.send(token);
    }
  );
});
```

Fig. 5. Server verifying the credentials of the application.

As the application calls the STT service with the voice input data, it sends a JSON response back to the front end. At the same time, the client-side script renders the sign-in page on the screen asking for the existing customer information to prevent unauthorized access. The new users can use the register page to register to the applications database. After the initial rendering is done and the components are mounted on the screen (sign-in form), users can use the STT service to enter their information to the sign-in or register form. The frontend will then send this information to the backend which in turn will be sent to the database for verification or registration purposes. After advancing through the initial check, the frontend will make a fetch request based on users' requests as shown in Fig. 6, which on success will return a promise that resolves to the response of that request as shown in Fig. 7 indicating keywords. These types of systems are basically used in the query-based applications where the systems return answers to user queries.

```
componentDidMount(){
  fetch('https://webmppcapstone.blob.core.windows.net/data/itemsdata.json')
  .then(response=>response.json())
  .then(data=>this.setState({data:data}))

}
```

Fig. 6. Fetch request.

In our use case scenario based on the keywords, the backend application returns a response containing a JSON dataset as shown in Fig. 8. The response contains all the product or merchandise related information like image links, names, costs, and descriptions of the products. It is then parsed and processed by the frontend application (see data flow in Fig. 4) to render the products on the screen. Users can then use their voice to activate the TTS service and give commands to search, add, or remove a particular item from the cart. In the case the searched item is not available, the TTS service is used to generate a voice response to inform the customers about it. So, by incorporating both Watson STT and TTS we can develop an intelligent application that can actually interact with the users and help users with accessibility problems.

```
{…}
  alternatives: (1) […]
    0: {…}
      confidence: 0.912
      timestamps: Array(3) [ (3) […], (3) […], (3) […] ]
      transcript: "Search for brownies. "
      <prototype>: Object { … }
    length: 1
    <prototype>: Array []
  final: true
  index: 0
  keywords_result: {…}
    Brownies: (1) […]
      0: Object { normalized_text: "brownies", start_time: 1.65, confidence: 0.989, … }
      length: 1
      <prototype>: Array []
    search: (1) […]
      0: Object { normalized_text: "search", start_time: 0.97, confidence: 1, … }
      length: 1
      <prototype>: Array []
    <prototype>: Object { … }
  <prototype>: Object { … }

{…}
  speaker_labels: (3) […]
    0: Object { from: 0.97, to: 1.37, speaker: 0, … }
    1: Object { from: 1.37, to: 1.65, speaker: 0, … }
    2: Object { from: 1.65, to: 2.42, speaker: 0, … }
    length: 3
    <prototype>: Array []
  <prototype>: Object { … }
```

Fig. 7. Screenshot of Response from Watson STT service.

```
(4) […]
  0: Object { category: "Household and Beauty", subcategories: (4) […] }
  1: Object { category: "Pantry Items", subcategories: (5) […] }
  2: Object { category: "Perishables", subcategories: (4) […] }
  3: Object { category: "Produce", subcategories: (2) […] }
  length: 4
  <prototype>: Array []
```

Fig. 8. The response about the products in JSON format.

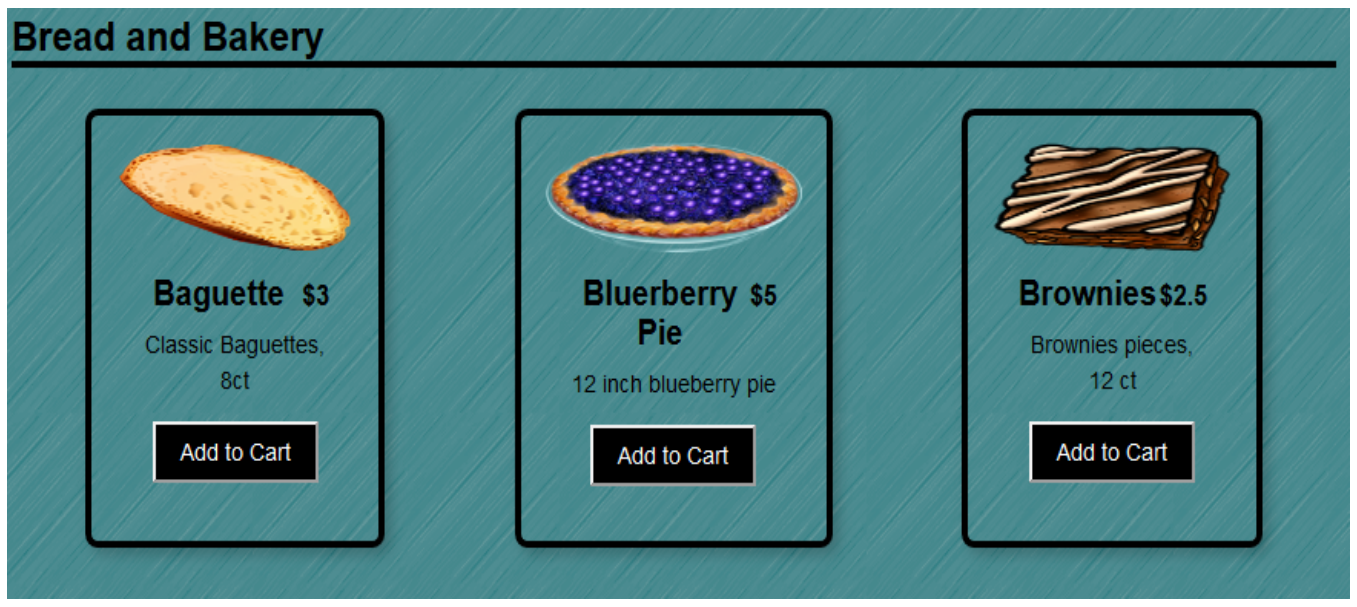

Fig. 9. Products page as shown on the website.

## *B. Ongoing Work*

The basic functionality of the system has been developed where users can use their voice commands with the application instead of typing to give inputs and perform a search to see available goods as shown in Fig. 9. The next step would be to incorporate the TTS service in the application so that in the case the application is not able to service users' commands, it can inform the same to the users using a generated voice, complete purchase for the user, and confirm or cancel transactions. This application will address the accessibility problems of the e-commerce websites to some extent. In the future, we will focus on incorporating a virtual assistant to provide recommendations. In this way, we will be able to implement a system that will enhance users' comfort and experience.

## VII. CONCLUSIONS

We present a study of some of the cutting edge SRSs in this research, a summary of performance comparisons of some of the popular SRSs from the literature, a taxonomy for categorizing SRSs based on their functionality, and a preliminary prototype of a voice-controlled e-commerce web application using IBM Watson STT and TTS services. Enhancing accessibility to commercial websites is crucial for today's e-commerce dependent economy. We implement a SRS-based e-commerce web application with a view to leverage accessibility to web applications for the visually-impaired users such that they can use their voice as a means to operate the application. SRS enabled applications can enhance usability for all users by promoting ease of interaction and multi-tasking, and support a lean environment where users can make requests using natural language.

For the future work, we plan to compare the performances of other cloud-based services such as Google and Amazon in the context of the same e-commerce application. The application can be extended to identify a specific speaker based on the voice. Furthermore, the application can also be equipped with the language translation APIs that will allow users to give commands to the application in their own native language. This can prove to be beneficial for the users who are restrained from using the websites because of the language barrier. The application can be designed in such a way, that it can render user-specific interfaces based on their interests. This means that the interface can be rendered distinctively for every user thus making the application more appealing and user-friendly. The application can also be extended to other domains such as for online education, providing services at government service kiosks, and for emergency online assistance.

## ACKNOWLEDGMENT

We like to acknowledge the support from IBM and the Centre for Advanced Computing at Queen's University to provide us access to the Watson services, and the online course website for the product data used in this research.

## REFERENCES

[1] National Research Council. (2002). Technology and Assessment: Thinking Ahead--Proceedings from a Workshop. National Academies Press.

[2] N. Indurkhya, F. J. Damerau, Handbook of natural language processing Vol. 2, CRC Press, 2010.

[3] D. Li and Y. Dong, "Deep learning: methods and applications," Foundations and Trends{\textregistered} in Signal Processing, vol. 7, pp. 197-387, 2014.

[4] B. Hashemian, "Analyzing web accessibility in Finnish higher education", *ACM SIGACCESS Accessibility and Computing*, no. 101, pp. 8-16, 2011.

[5] K. Nahon, I. Benbasat and C. Grange, "The Missing Link: Intention to Produce Online Content Accessible to People with Disabilities by Non-professionals," *2012 45th Hawaii International Conference on System Sciences*, Maui, HI, 2012, pp. 1747-1757.

[6] W3C, "Introduction to web accessibility", available from: https://www.w3.org/TR/2018/REC-WCAG21-20180605/

[7] M. Akram and R. Bt Sulaiman, "A Systematic Literature Review to Determine the Web Accessibility Issues in Saudi Arabian University and Government Websites for Disable People", *International Journal of Advanced Computer Science and Applications*, vol. 8, no. 6, 2017.

[8] M. A. Anusuya and S. K. Katti, "Speech Recognition by Machine," *International Journal of Computer Science and Information Security, IJCSIS,* vol. 6, no. 3, pp. 181-205, 2010.

[9] K. Christian, B. Kules, B. Shneiderman and A. Youssef, "A Comparison of Voice Controlled and Mouse Controlled Web Browsing," Proceedings of the Fourth International ACM Conference on Assistive Technologies, no. ACM, pp. 72-79, 2000.

[10] A. B. BAJPEI, M. S. SHAIKH and N. S. RATATE, "VOICE OPERATED WEB BROWSER," *International Journal of Soft Computing and Artificial Intelligence,* vol. 3, no. 1, pp. 30-32, May-2015.

[11] S. Han, G. Jung, B.U. C. Minsoo Ryu and J. Cha, "A Voice-controlled Web Browser to Navigate Hierarchical Hidden Menus of Web Pages in a Smart-tv Environment," Proceedings of the 23rd International Conference on World Wide Web, pp. 587-590, 2014.

[12] S. Sagar, V. Awasthi, S. Rastogi, T. Garg, S. Kuzhalvaimozhi, "Web application for voice operated e-mail exchange",

[13] V. K{\"e}puska and B. Gamal, "Comparing speech recognition systems (Microsoft API, Google API and CMU Sphinx," *Journal of Engineering Research and Application,* vol. 7, no. 3, pp. 20-24, 2017.

[14] C. Gaida, P. Lange, R. Petrick, P. Proba, A. Malatawy and D. Suendermann-Oeft, "Comparing open-source speech recognition toolkits," *Tech. Rep., DHBW Stuttgart,* 2014.

[15] E. Gazetić, "Comparison Between Cloud-based and Offline Speech Recognition Systems". Master's thesis. Technical University of Munich, Munich, Germany, 2017.

[16] M. Alshamari, "Accessibility evaluation of Arabic e-commerce web sites using automated tools," Journal of Software Engineering and Applications, vol. 9, no. 09, p. 439, 2016.

[17] C. McNair, "Worldwide retail and ecommerce sales: emarketer's updated forecast and new mcommerce estimates for 2016–2021," Industry Report, eMarketing , 2018.

[18] Y. Borodin, J. P. Bigham, G. Dausch and I. Ramakrishnan, "More than meets the eye: a survey of screen-reader browsing strategies," Proceedings of the 2010 International Cross Disciplinary Conference on Web Accessibility (W4A), p. 13, 2010.

[19] Alison DeNisco Rayome, 2017, "Why IBM's speech recognition breakthrough matters for AI and IoT", white paper, online at: https://www.techrepublic.com/article/why-ibms-speech-recognition-breakthrough-matters-for-ai-and-iot/